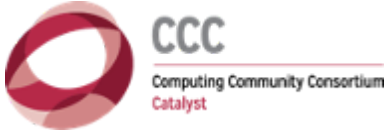

# A Research Ecosystem for Secure Computing

*A Computing Community Consortium (CCC) Quadrennial Paper*

*Nadya Bliss (Arizona State University), Lawrence A. Gordon (University of Maryland), Daniel Lopresti (Lehigh University), Fred Schneider (Cornell University), and Suresh Venkatasubramanian (University of Utah)*

Computing devices are vital to all areas of modern life and permeate every aspect of our society. The ubiquity of computing and our reliance on it has been accelerated and amplified by the COVID-19 pandemic. From education to work environments to healthcare to defense to entertainment – it is hard to imagine a segment of modern life that is not touched by computing.

The security of computers, systems, and applications has been an active area of research in computer science for decades. However, with the confluence of both the scale of interconnected systems and increased adoption of artificial intelligence, there are many research challenges the community must face so that our society can continue to benefit and risks are minimized, not multiplied. Those challenges range from security and trust of the information ecosystem (the subject of a separate white paper) to adversarial artificial intelligence and machine learning.

Along with basic research challenges, more often than not, securing a system happens after the design or even deployment, meaning the security community is routinely playing catch-up and attempting to patch vulnerabilities that could be exploited any minute. While security measures such as encryption and authentication have been widely adopted, questions of security tend to be secondary to application capability. There needs to be a sea-change in the way we approach this critically important aspect of the problem: new incentives and education are at the core of this change.

As more devices become connected and join the Internet of Things, the importance of prioritizing security *alongside* capability only increases. Now is the time **to refocus research community efforts on developing interconnected technologies with security "baked in by design" and creating an ecosystem that ensures adoption of promising research developments**.

To realize this vision, two additional elements of the ecosystem are necessary – **proper incentive structures for adoption** and **an educated citizenry that is well versed in vulnerabilities and risks**. The research, design, and development incentive structures need to shift so that the first question asked when designing a new piece of connected technology is not just *'does it work?'*, but also *'is it secure?'*.

As outlined, this is a tightly coupled ecosystem that requires both focus and investment in each of these goals and in interactions between them. Below, we outline some potential initiatives and topics in each of the four focus areas: *research, transition and adoption, incentives, education*.

### (Computing) Research

Sustained investment in computing security research is vital for all security of diverse application domains as well as economic security. As new technologies, such as artificial intelligence and quantum computing, are developed, it is important to consider how both to leverage those technologies for increased security and potential emergence of novel vulnerabilities when those new technologies are deployed.

A few timely research topics include but are not limited to:

- **Artificial intelligence (AI) and security**:
    - Vulnerabilities of AI systems particularly in cyber-physical systems such as autonomous vehicles;
    - Leverage of AI techniques in context of cyber defense;
- **Quantum computing and security**:
    - Advances in quantum computing have the potential to jeopardize the security of current encryption systems, so sustained research in new quantum cryptographic approaches and more generally the impact of quantum computing on what we think of as "security" today are important;
- **Internet of Things (IoT) (a.k.a., cyber-physical systems) and next generation communication systems:**
    - Vulnerabilities, security, and resilience of IoT devices, which are rapidly proliferating, require new approaches to security in constraint communication and/or processing environments, including novel communication technologies such as 5G (the subject of a separate white paper);
- **Hardware security**:
    - Computer architectures designed for security at the hardware layer and resilient to recently deployed hardware attacks (Meltdown and Spectre);
- **Critical infrastructure security**:
    - Critical infrastructure such as electrical grids and other control systems remain highly vulnerable and have different operational requirements;
- **Risk assessment of centralized architectures**:
    - As more of the economy moves to centralized cloud architectures, research in the areas of risks of these single points of failure and associated mitigation are critical.

A number of these topics are already being invested in by agencies such as the Defense Advanced Research Projects Agency (DARPA) and the National Science Foundation (NSF), however as investments in AI and quantum computing research continue to scale it is vital to appropriately co-invest in emerging security challenges associated with these new capabilities in parallel to avoid playing catch up after the fact. While for the last few decades computer security research has been advancing alongside capability research, we recommend creating consistent and ongoing touchpoints and co-design opportunities specifically in the new technology areas (AI and quantum).

As computing touches more and more application and mission sectors, it is important to prioritize research investments in computer security in sector specific agencies, such as increasing investments in Department of Energy (specifically, around supervisory control and data acquisition (SCADA) systems and electrical grid security), Department of Transportation (around autonomous systems and smart infrastructure), and National Institutes of Health (around security of telehealth and cybersecurity of biological research systems).

### Transition and Adoption

Transition and adoption of new technology is a well-documented challenge. In the context of security, however, the lack of transition of cyber security research has made us vulnerable as a nation. With the accelerated adoption of computing capability, it is timely to consider examples of effective structures for adoption. FFRDCs (Federally Funded Research and Development Centers) and UARCs (University Affiliated Research Centers) are examples of organizational constructs that create environments to de-risk cutting edge technology in context of mission specific applications (most commonly, defense applications).

Given the way computing has permeated our society, it would be beneficial to consider both persistent/longer term investments (leveraging existing UARC's and FFRDC's and/or creation of new ones) and shorter term (accelerator-like) constructs that would de-risk and aid transition of cyber security research for various sectors of the economy. These constructs could be specific to a technology layer (for example, hardware security) or to an application domain (for example, healthcare).

### Incentive Structures

As suggested earlier, as new efficiency in communication systems evolves and more and more domains move online, new research is necessary in securing computer architectures, securing communication networks, and mitigating and repairing vulnerabilities. This is an important area to continue focusing on, working towards foundational advances.

Developing completely secure systems is not likely to ever be attainable – new technologies, new applications, and new system paradigms create new attack surfaces that can be exploited. This is inevitable. And it is difficult to design regulation and create incentives without being able to assess risk effectively. How would a policy maker incentivize a secure system design, if there is no clear definition of what that means?

Therefore, more focused efforts on measuring the security of interconnected systems will be of significant benefit. A key challenge in building a secure system is that it is easier to assess capability than it is to assess security – we can easily check if the system does the task it was designed to do or that it does so within the computational and time constraints specified, while it is much harder, if not impossible, to conduct an exhaustive assessment of all of the vulnerabilities of a system. This becomes even more complex when that system must be assessed in the context of other, interacting systems.

A means to address this challenge could be through the creation of a large-scale interdisciplinary research program to quantitatively study incentive structures for secure computing; security baked-in by design. It is vital that this initiative be structured as a public/private partnership, as it would be fundamentally impossible to cause change without buy-in from the private sector. A strong engagement from the national security community here would also be highly beneficial because, in the mission focused domain of national security, system capabilities and system vulnerabilities are typically considered in concert.

Additionally, there would be significant benefits in creating a large-scale initiative to assess system security and develop not just metrics (is it secure?), but also approaches to evaluate potential security risks (if it is not secure, what are the costs of that?) that could be achieved through combinations of technologies.

Determining the right amount to invest in cybersecurity is an important concern for all organizations. A fundamental issue associated with this concern is that profit-oriented firms focus on the private costs associated with cybersecurity breaches, but routinely ignore the externalities associated with the breaches (i.e., the social costs of breaches are routinely not considered in cybersecurity investment decisions by profit-oriented firms). It would be extremely beneficial for the federal government to fund research that is directed toward developing economic incentives that help to align a firm's cybersecurity investment decisions with the socially optimal cybersecurity investment decisions (i.e., the decisions that take into account externalities, as well as the organization's private costs).

These initiatives should build upon existing standards developed by NIST and other institutions. Assessment of metrics and incentives should be coupled to policy recommendations in an ongoing fashion, as both technologies and policies evolve.

### Education, Training, and Workforce

As computing is being used by broad segments of the population, investment in education with a focus on a security mindset is critical. This is important not only in computer science degree programs, but elsewhere throughout higher education and more broadly across the entire population. There has been significant progress on increasing technical literacy in K-12 and broadening access to certificate and college-level computer science curriculum. However, much of the education in risks and assessment of threats and development of secure systems is still lacking. Increasing security risk elements in the K-12 curriculum, adding a focus on designing for security for computer science degrees (building on NSA/DHS cybersecurity academic excellence certifications), and adding security and vulnerability components to interdisciplinary computational curriculums should be the priorities.

**Summary of Recommendations**

As both the scale and complexity of computing systems in all aspects of our lives continue to grow, it is vital to consider holistically what is necessary not just to manage existing risks, but to plan and mitigate future ones. This prioritization of security is needed for research, transition, incentive, and education initiatives. Attention should be paid not just to each domain individually, but to the interactions between them.

- Sustained investment in computer science research across both basic science and mission focused agencies;
- Creation of mission/sector focused accelerators to support transition of relevant cybersecurity research into application and industry;
- A multi-disciplinary effort and public/private partnership around metrics and incentives for security with a goal of continuously producing policy recommendations;
- Investment in lifelong learning and training to support a "security mindset" across the entire population.

*This white paper is part of a series of papers compiled every four years by the CCC Council and members of the computing research community to inform policymakers, community members and the public on important research opportunities in areas of national priority. The topics chosen represent areas of pressing national need spanning various subdisciplines of the computing research field. The white papers attempt to portray a comprehensive picture of the computing research field detailing potential research directions, challenges and recommendations.*

*This material is based upon work supported by the National Science Foundation under Grant No. 1734706. Any opinions, findings, and conclusions or recommendations expressed in this material are those of the authors and do not necessarily reflect the views of the National Science Foundation.*

*For citation use: Bliss N., Gordon L., Lopresti D., Schneider F., & Venkatasubramanian S. (2020) A Research Ecosystem for Secure Computing.*
*https://cra.org/ccc/resources/ccc-led-whitepapers/#2020-quadrennial-papers*